\documentclass[journal]{IEEEtran}
\usepackage{cite}
\usepackage{array}

\usepackage{amssymb}
\usepackage{amsthm, amsmath, amsfonts, amssymb}
\usepackage{mathtools}
\usepackage{algorithm}
\usepackage{algpseudocode}
\usepackage{graphicx}
\usepackage{textcomp}
\usepackage{subfig}  
\usepackage{xcolor}
\usepackage{float}
\usepackage{multirow}

\graphicspath{{figures/}}

\newcommand{\R}{\mathbb R}

\newcommand{\mrm}[1]{\mathrm{#1}}
\newcommand{\bs}[1]{\boldsymbol{#1}}
\newcommand{\diff}{\ensuremath{\mathrm{d}}}

\begin{document}
\title{State and parameter estimation for retinal laser treatment}

\author{Viktoria Kleyman, Manuel Schaller, Mario Mordmüller, Mitsuru Wilson, Ralf Brinkmann, Karl Worthmann and Matthias A.\ Müller
\thanks{The collaborative project "Temperature controlled retinal laser
	treatment" is funded by the German Research Foundation (DFG)
	under the project number 430154635 (MU 3929/3-1, WO 2056/7-1,
	BR 1349/6-1). K. Worthmann gratefully acknowledges funding
	by the German Research Foundation (DFG; grant WO 2056/6-1,
	project number 406141926)}
\thanks{V. Kleyman and M. A. Müller are with the Leibniz University Hannover, Institute of Automatic Control, Germany (e-mail: \{kleyman, mueller\}@irt.uni-hannover.de). M. Schaller, M. Wilson and K. Worthmann are with the Technische Universität Ilmemau, Institute of Mathematics, Germany (e-mail:\{manuel.schaller, mitsuru.wilson, karl.worthmann\}@tu-ilmenau.de). R. Brinkmann and M. Mordmüller are with the University of Lübeck, Institute of Biomedical Optics, (e-mail:\{m.mordmueller, ralf.brinkmann\}@uni-luebeck.de).}}

\maketitle

\begin{abstract}
Adequate therapeutic retinal laser irradiation needs to be adapted to the local absorption. This leads to time-consuming treatments as the laser power needs to be successively adjusted to avoid under- and overtreatment caused by too low or too high temperatures. Closed-loop control can overcome this burden by means of temperature measurements. To allow for model predictive control schemes, the current state and the spot-dependent absorption need to be estimated.
In this paper, we thoroughly compare moving horizon estimator (MHE) and extended Kalman filter (EKF) designs for joint state and parameter estimation. We consider two different scenarios, the estimation of one or two unknown absorption coefficients. For one unknown parameter, both estimators perform very similar. For two unknown parameters, we found that the MHE benefits from active parameter constraints at the beginning of the estimation, whereas after a settling time both estimators perform again very similar as long as the parameters are inside the considered parameter bounds.     
\end{abstract}


\section{Introduction}

 Laser photocoagulation is a widely used application to treat retinal diseases such as diabetic retinopathy and macula edema. However, the treatment must be individually adapted by the ophthalmologist to each patient and each irradiation site due to varying parameters such as the absorption within the eye. Since the treatment time is in the range of a few hundred milliseconds, the ophthalmologist cannot intervene during the irradiation process. Only after the treatment at one spot, the laser power can be adjusted for the next spot. This procedure is time-consuming and provides only limited prevention against over- or undertreatment. In addition, recent studies show that also hyperthermia can be beneficial for diseases such as diabetic macula edema \cite{Luttrull.2012,Iwami.2014, Lavinsky.2016}. During hyperthermia treatments, no tissue is destroyed and the target spot at the retina remains invisible. Thus, the physician can not determine whether the laser power has been appropriately selected. For this, it is necessary that the physician obtains additional information, such as the temperature at the irradiated spot.

A method for measuring a depth-weighted volume temperature was developed in \cite{Brinkmann.2012}. To this end, a piezo-electric transducer is integrated into a commercially available contact lens. A repetitively pulsed laser beam induces pressure waves, which can be detected by the transducer. The temperature-dependent pressure wave amplitude can then be used to calculate the volume temperature. This temperature feedback enables the development of control strategies that can be used for both coagulation and hyperthermia applications. However, the volume temperature is only suitable for control to a limited extent, since the peak temperature, i.e., the highest temperature inside the tissue, can already cause damage before the volume temperature is above a (coagulation) threshold and is therefore crucial for a successful treatment.

In recent years, some first strategies for controlling the peak temperature were developed in \cite{Abbas.2020,Herzog.2018,Baade.2017}. As the peak temperature that occurs in the retinal pigment epithelium (RPE) as the strongest absorber cannot be measured directly, an offline determined function is used to convert the volume temperature into the peak temperature. To this end, constant laser powers and a constant ratio between the absorption coefficients within the tissue were assumed. Based on the approximation of the peak temperature, an open-loop control strategy was developed in \cite{Baade.2017} and first closed-loop (robust PID) schemes were presented in \cite{Abbas.2020} and \cite{Herzog.2018}. However, the approximated conversion to the peak temperature is only valid under certain assumptions, e.g., a constant laser power that does not hold in closed-loop.  
Furthermore, these schemes lack the ability to incorporate upper bounds on the peak temperature necessary for a safe treatment or a maximum laser power in the control design. In model predictive control (MPC) schemes, these safety-related constraints can be considered and their compliance can be guaranteed. We aim for an MPC scheme that incorporates a reduced order model that captures the dynamics of the (infinite-dimensional) heat diffusion equation. 
To this end, a good estimate of the state and parameters is crucial, which is the subject of this paper.  

In our previous work \cite{Kleyman2020}, we used a parametric model order reduction technique from \cite{Baur11} and extended it to polynomial parameter dependencies to obtain a parametric reduced order model that allows for estimation of the absorption coefficients after model reduction. In \cite{Schaller.2022a}, we performed a thorough case study using 250 measurement spots from 25 porcine eyes in order to determine the range for the absorption coefficients located between the photoreceptors of the neural retina and the choroid (cf. Fig.~\ref{fig:cylinders}). Moreover, in \cite{Schaller.2022a} we examined an alternative model reduction technique applying a global basis approach~\cite{Benner.2015} in combination with a discrete empirical interpolation method~\cite{Chaturantabut2010}, which was shown to reduce the model order reduction error compared to the method previously proposed in \cite{Kleyman2020}. 

In this paper, we develop extended Kalman filters and moving horizon estimators in order to estimate the state and spot-dependent absorption coefficients in a joint fashion. Since in general, the extended Kalman filter (EKF) can fail due to, e.g., poor guesses of the initial conditions and tuning parameters as shown in \cite{Haseltine.2005}, we perform a thorough comparison of both approaches to evaluate which estimation technique is suited in the context of retinal laser treatment. We consider two different scenarios: first, a simpler one where the absorption coefficient in the choroid is assumed to be constant and only the absorption coefficient in the RPE is estimated.
This is motivated by our case study \cite{Schaller.2022a}, which showed that the input-output sensitivity is higher with respect to the RPE absorption coefficient than to the choroid absorption coefficient. 
In a second scenario, both absorption coefficients of the RPE and the choroid are estimated independently. We investigate both estimation techniques in simulation and with real measurement data from porcine eyes in 1\,kHz. In the first scenario, both estimators show very similar behavior. In the second scenario, the moving horizon estimator (MHE) profits from active parameter constraints at the beginning (of the estimation) which could be crucial in closed-loop estimation. After a settling time, both estimators perform similar if the parameters are inside the considered bounds in the MHE formulation. In the rare but possible case that the parameters are outside the bounds, MHE cannot estimate the correct values and, therefore, EKF outperforms MHE. Despite the differences in the estimation of two parameters, we found that both estimators are well suited with regard to our application.    
The results of this work hence pave the way for successful application of model-based control approaches, in particular model predictive control. A first proof-of-concept in this direction was recently shown by us in \cite{Mordmueller.2021,Schaller.2022b}, where we use MPC in combination with EKF.

Very recently, some related work has been presented in \cite{Arnold.2022}, where the absorption coefficient in a homogeneous tissue is estimated via an ensemble Kalman filter (EnKF). In contrast, in this work we consider an inhomogeneous tissue with different absorption coefficients; furthermore, since we aim for (computationally demanding) model predictive control, state and parameter estimation has to be real-time capable and hence cannot be done using a large-scale full order model as in \cite{Arnold.2022}. Instead, we show that EKF and MHE schemes can be developed based on a suitably defined reduced order model.

A preliminary version of parts of this paper has appeared in the conference proceedings \cite{Kleyman.2021a}, where a first simple EKF and MHE design was proposed for the case of one parameter (RPE absorption coefficient to be estimated) based on the reduced order model from \cite{Kleyman2020}. Compared to \cite{Kleyman.2021a}, in this work we (i) use the better suited MOR technique proposed in \cite{Schaller.2022a}, (ii) provide a much more comprehensive simulation case study, (iii) evaluate the performance of both estimators using real measurement data from porcine eyes in detail, and (iv) develop estimation schemes in case that both the RPE and choroid absorption coefficients are estimated independently.

This paper is organized as follows. In Section~\ref{sec:setup}, we present our experimental setup. In Section~\ref{sec:modeling}, we introduce our spatially discretized model that is described by an underlying partial differential equation and give insights to the case study and the model order reduction technique that serve as a foundation for our observer design and evaluation. Then, in Section~\ref{sec:observerdesign}, we present both MHE and EKF designs for joint state and parameter estimation before we compare both observers in the case of one unknown absorption coefficient in simulation and with measurement data in Section~\ref{sec:Comparison}. Afterwards, we take a closer look to the estimation with two unknown absorption coefficients (in RPE and choroid) in Section~\ref{sec:2p}. Last, in Section~\ref{sec:conclusion}, we draw conclusions and give an outlook to future work.

\section{Experimental Setup}\label{sec:setup}
\begin{figure}
	\centering
	\includegraphics[width = 0.5\textwidth]{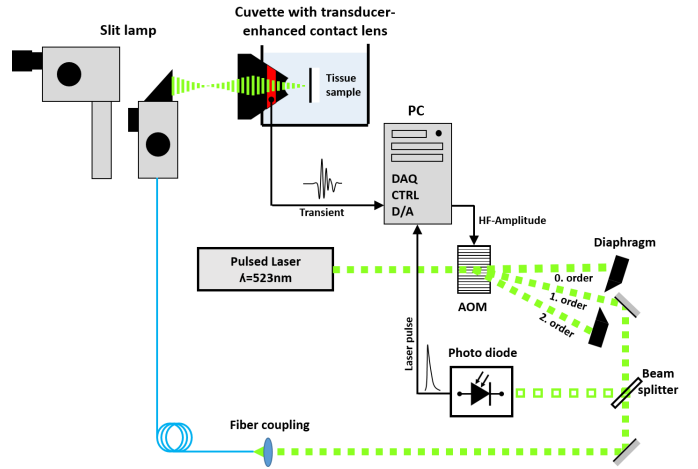}
	\caption{Sketch of the experimental setup.}
	\label{fig:setup}
\end{figure} 

\noindent Figure \ref{fig:setup} depicts the experimental setup. The upper left part shows an ophthalmic slit lamp with integrated laser link. The laser beam is focused onto the tissue sample by means of an ophthalmic contact lens. The contact lens was customized with a ring-shaped piezo-ceramic transducer to detect the pressure wave induced by the laser pulse. The contact lens is attached to a sample cuvette filled with sodium chloride solution of $0.9\,\%$.
The beam of a Q-switched and frequency-doubled solid-state Nd:YLF laser with a wavelength of $\lambda=523\, \text{nm}$ is aimed through an acousto-optic modulator (AOM). Upon ultrasonic modulation of the AOM, the laser beam is split into different orders of diffraction. Here, only the first-order diffraction is passed through a diaphragm and used for sample irradiation. By varying the ultrasonic amplitude the diffraction efficiency can be controlled with frequencies higher than 10\,kHz. The beam is coupled to an optical fiber and guided to the slit lamp, and then applied to the tissue sample. The spot diameter on the tissue sample is $D=200\; \mu\text{m}$. Both, pressure transient and laser pulse signals are recorded by a fast data acquisition board and processed with C/C++ MFC software. 
The laser is operated with a pulse repetition rate of $10 \; \text{kHz}$. Every 10th pulse is set to a fixed probe energy and used for temperature measurement. All experiments are conducted on explants of enucleated porcine eyes with removed retina. 

More details about the optical setup and measurement methodology can be found in \cite{Mordmueller.2021}.

\section{Modeling and Model Order Reduction}\label{sec:modeling}
\begin{figure}
	\centering  
	\resizebox{0.75\columnwidth}{!}{\includegraphics{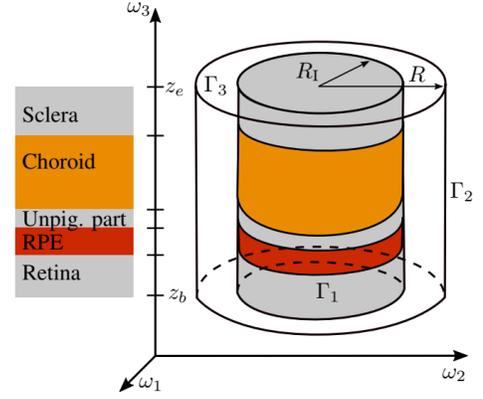}}
	\caption{Schematic illustration of the inner and outer cylinder with the five considered layers of the eye fundus.}
	\label{fig:cylinders}
\end{figure}

\noindent In the following, we present our model describing the evolution of the temperature distribution in time and space that is induced by laser irradiation. 
The computational domain is composed by five different layers of the eye fundus, namely the retina, the retinal pigment epithelium (RPE), the unpigmented part (Bruch's membrane), the choroid, and the sclera, compare Fig.~\ref{fig:cylinders}. We consider an inner cylinder with the radius ${R_\mrm{I}}$ of the irradiated spot. As the heat diffuses in the surrounding tissue, we consider a second, outer cylinder in the domain $\Omega\subset\mathbb{R}^3$ that allows for Dirichlet boundary conditions if $\Omega$ is large enough. The difference between the ambient temperature and the tissue is given by $T(\omega,t)$ at spatial coordinate $\omega$ and time $t$. The time evolution of this temperature difference is modeled by the heat diffusion equation
\begin{align}
	\rho C_\mrm{p} \frac{\partial T(\omega,t)}{\partial t}-k\Delta T(\omega,t)=Q(\omega,t)\; \forall\,  
	(\omega,t)\in\Omega\times(0,t_\mrm{f})
	\label{eq:pde}
\end{align}
with boundary and initial conditions 
\begin{align}
	\begin{split}
		T(\omega,t)&=0\quad \text{$\forall\, (\omega,t)\in\Gamma\times (0,t_\mrm{f})$},\\ 
		T(\omega,0)&=0 \quad \forall\, \omega\in\Omega,
	\end{split}
	\label{eq:BC}
\end{align}
where $\Gamma:=\Gamma_1\cup\Gamma_2\cup\Gamma_3$ are the boundaries of the outer cylinder.
The heat capacity $C_\mrm{p}$, the thermal conductivity $k$ and the density $\rho$ are assumed to be constant and those of water, see \cite{Kleyman.2021a}.
The light-tissue interaction is modeled as a heat source $Q(\omega,t)$ in the inner cylinder given by Lambert-Beer law:
\begin{align}
\label{eq:heat_source}
Q(\omega,t):=
\begin{cases}\frac{u(t)}{\pi R_\mrm{I}^2}\mu(\omega_3) e^{-\int_0^{\omega_3}\mu(\zeta)\mrm{d}\,\zeta}, &\text{if}\, \omega_1^2+\omega_2^2\leq R_\mrm{I}^2,\\ 
0,&\text{otherwise,}\end{cases}
\end{align}
where the input $u(t)$ is the laser power and $\mu$ is the absorption coefficient in the corresponding layer. The absorption of the laser light with a wavelength of 523\,nm mainly takes place in the RPE and choroid, hence we consider absorption in these two layers with the coefficients $\mu_\mrm{rpe}$ and $\mu_\mrm{ch}$ for the RPE and the choroid, i.e.,
\begin{align*}
\mu(\omega_3) = \begin{cases}
\mu_\mrm{rpe}, \quad &\text{if } \omega_3 \in [z_1,z_2],\\
\mu_\mrm{ch}, \quad &\text{if } \omega_3 \in [z_3,z_4],\\
0, &\text{otherwise},
\end{cases}
\end{align*}
with $z_i \text{ and } i = 1,...,4$ according to Fig.~\ref{fig:cylinders}.

The measurable output of the system, that is the volume temperature $T_\mrm{vol}(t)$, can be expressed as the depth-weighted integral of all temperatures in the irradiated inner cylinder     
\begin{align}
	T_\mrm{vol}(t)= \int_{z_\mrm{b}}^{z_\mrm{e}} T_\mrm{mean}(t,\omega_3) \mu(\omega_3)e^{-\int_{0}^{z}\mu(\zeta)\diff \zeta} \mrm{d} \omega_3,
	\label{eq:Tvol}
\end{align}
where $z_\mrm{e}-z_\mrm{b}$ is the length of the cylinder. The temperature $T_\mrm{mean}(t,\omega_3)$ is the mean of all temperatures in the irradiated ($\omega_1$, $\omega_2$)-plane at $\omega_3$. Using cylindrical coordinates $(r,\phi,z)$, this mean temperature is then given by
\begin{align*}
T_\mrm{mean}(t,z) &=\frac{1}{\pi R_\mrm{I}^2}\int_{0}^{2\pi} \mrm{d}\,\phi \int_{0}^{R_\mrm{I}} r x(r,z,t)\mrm{d}\,r.
\end{align*}    
Our second output is the peak temperature $T_\mrm{peak}$ that is crucial for safe and effective treatment. Numerical simulations show that during heating the hottest point lies in the center of the RPE layer. We emphasize that the peak temperature is not a measurable output but the quantity that has to be controlled. Hence, both outputs are considered for model order reduction to obtain a suitable model for estimation and control.
 
We perform a spatial discretization via finite differences to obtain a parameter-dependent finite-dimensional state-space model of order $n_\mrm{f}$:
\begin{align}
	\label{eq:fullsys_c}
	\begin{split}
		\dot x^\mrm{f}(t)&=A^\mrm{f}x^\mrm{f}(t)+b^\mrm{f}(\mu)u(t),\quad x^\mrm{f}(0)=0\\
		\quad y(t)&= C^\mrm{f}(\mu)x^\mrm{f}(t) = \begin{pmatrix}	c_\mrm{vol}^\mrm{f}(\mu)\\ c_\mrm{peak}^\mrm{f}\end{pmatrix}x^\mrm{f}(t) ,\quad t\geq 0,
	\end{split}
\end{align}
where $ \mu=(\mu_{rpe}, \mu_{ch})^\top$, $A^\mrm{f}\in \R^{n_\mrm{f} \times n_\mrm{f}}$, $x^\mrm{f} \in \R^{n_\mrm{f}}$, 
$b^\mrm{f} \in \R^{n_\mrm{f}}$, $C^\mrm{f}(\mu)\in \R^{2 \times n_\mrm{f}}$, and $\quad y(t)= (T_\mrm{vol}(t),\, T_\mrm{peak}(t))^\top$. 

For more details regarding the modeling, we refer the reader to our previous work \cite{Kleyman2020}.

\textbf{Parametrization.} In order to avoid scaling issues in the parameter estimation techniques, we parameterize the absorption coefficients with respect to baseline values $\mu_\mrm{0,rpe}$ respectively\ $\mu_\mrm{0,ch}$ of the literature \cite{Brinkmann.2012}. To this end, we set
\begin{align}
    \mu_\mrm{rpe} = \alpha_\mrm{rpe} \mu_\mrm{0,rpe},\qquad \mu_\mrm{ch} = \alpha_\mrm{ch} \mu_\mrm{0,ch}.
\end{align}
for positive prefactors $\alpha_\mrm{rpe}$ and $\alpha_\mrm{ch}$ that will be estimated.
In our previous work \cite{Schaller.2022a}, we performed a parameter study with 250 treatment spots and obtained the empirical means $\bar{\alpha}_\text{rpe}=0.7636$ and $\bar{\alpha}_\text{ch} = 0.0986$ and the empirical standard deviations $\sigma_\text{rpe} = 0.1907$ and \ $\sigma_\text{rpe} = 0.0281$, respectively. Thus, the parameter domain that was used for parametric model reduction and that will be used in the MHE in Subsec.~\ref{sec:MHE} is set to
\begin{align}
\label{eq:paramdom}
\begin{split}
       	    \mathcal{D} & = [\bar\alpha_{\text{RPE}}-2\sigma_{\text{RPE}},\bar\alpha_{\text{RPE}}+2\sigma_{\text{RPE}}]\\
	    &\qquad\qquad\qquad\qquad\times[\bar\alpha_{\text{ch}}-2\sigma_{\text{ch}}, \bar\alpha_{\text{ch}}+2\sigma_{\text{ch}}] \\ & = [0.3822,1.1451]\times[0.0424,0.1548].
\end{split}
\end{align}

\begin{figure}
	\centering
	\includegraphics{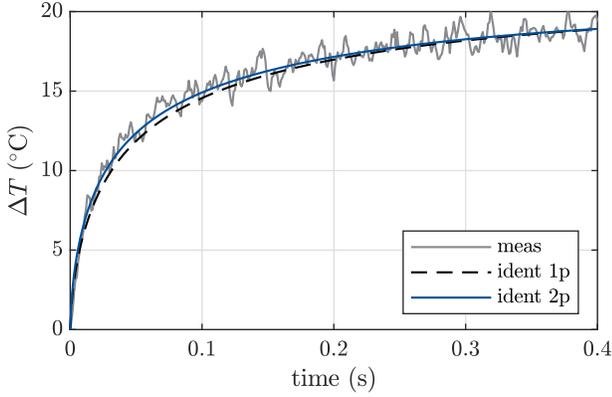}
	\caption{Difference between 1p and 2p offline identification}
	\label{fig:meas11_Tvol2p}
\end{figure}

Moreover, a theoretical and numerical sensitivity analysis both in time and frequency domain revealed that the sensitivity w.r.t.\ the RPE absorption coefficient is higher w.r.t.\ its counterpart in the choroid. Hence, we will first investigate the case where we fix the absorption coefficient in the choroid to its empirical mean and only estimate the absorption in the choroid. This simplification, however, leads to a different temperature increase, 
cf.\ Fig.~\ref{fig:meas11_Tvol2p}, motivating the inspection of both cases of one and two parameters. The former will be considered as the 1p-case with $\alpha = \alpha_\text{rpe}$ in Section~\ref{sec:Comparison} and the latter as the 2p-case with $\alpha = (\alpha_\text{rpe},\alpha_\text{ch})$ in Sec.~\ref{sec:2p}.

\textbf{Parametric Model Order Reduction}  In order to enable real time estimation and control of the parametric model~\eqref{eq:fullsys_c}, we apply parametric model order reduction (pMOR) to system~\eqref{eq:fullsys_c}. 
For the model order reduction, we use a common approach of employing a global basis that is, e.g., described in \cite{Benner.2015}. The global basis matrices $W^\top \in \R^{n \times n_\mrm{f}}$ and $V \in \R^{n_\mrm{f} \times n}$ are obtained from concatenating local (parameter-dependent) basis matrices evaluated for several parameter samples. This concatenation step is followed by a singular value decomposition (SVD) to avoid rank-deficient matrices $W^\top$ and $V$. 
The local bases are computed by means of the \textit{iterative rational Krylov algorithm} (IRKA) from \cite{GugeAnto08}. We used the sssMOR MATLAB package, see \cite{MORtool}, for the computation of the local bases. 

The evaluations of the parameter-dependent high dimensional nonlinearities $b^\mrm{f}(\alpha)$ and $c_\mathrm{vol}^\mrm{f}(\alpha)$ after model reduction, i.e., $W^\top b^\mrm{f}(\alpha)$ and $c_\mathrm{vol}^\mrm{f}(\alpha)V$ are computationally expensive during estimation (due to the parameter dependency). Hence, we use the discrete empirical interpolation method (DEIM) proposed in \cite{Chaturantabut2010} to approximate these nonlinearities by lower-dimensional surrogates.
DEIM selects interpolation indices that are used to project an approximation of the nonlinearity into a lower-dimensional subspace. The input vector $b^\text{f}(\alpha)$ can be approximated by
\begin{align}
	b^\text{f}(\alpha) \approx U_\mrm{b}(P_\mrm{b}^\top U_\mrm{b})^{-1} \tilde{b}(\alpha),
	\label{eq:DEIM_B}
\end{align}
where $P_\mrm{b} \in \R^{n_\mrm{f}\times d}$ is a matrix containing $d$ columns of the identity $I_{\mrm{n_\mrm{f}}}$, $\tilde{b}(\alpha) = P_\mrm{b}^\top b^\text{f}(\alpha)$ and the subspace $U_\mrm{b} \in \R^{n_\mrm{f}\times d}$. The computation of these matrices is described in detail in \cite{Chaturantabut2010}. The nonlinear output operator $c_\text{vol}^\text{f}(\alpha)$ can be approximated in the same fashion, i.e., $\tilde{c}(\alpha) = c_\text{vol}^\text{f}(\alpha) P_\mrm{c}$, $P_\mrm{c} \in \R^{d \times n_\mrm{f}}$. Note that the computations of $P_{\{\mrm{b,c}\}}$ and $U_{\{\mrm{b,c}\}}$ are independent of the subsequently applied model order reduction technique. Only the low dimensional $\tilde{b}(\alpha) \in \R^d$ and $\tilde{c}(\alpha) \in \R^d$ have to be evaluated for certain $\alpha$. This enables an efficient computation of the parameter-dependent reduced order model. 

Combining DEIM with the global basis approach was successfully applied in our previous work \cite{Schaller.2022a} and leads to a reduced system of dimension $n\ll n^\mrm{f}$:
\begin{align}
	\begin{split}
		\dot x(t)&=A x(t)+b(\alpha)u(t),\\
		y(t)&=C(\alpha)x(t),
	\end{split}
	\label{eq:red_c}
\end{align}
with $A = (W^\top V)^{-1}W^\top A^\mrm{f}V$, \\ $b(\alpha) = (W^\top V)^{-1} W^\top U_\mrm{b} (P_\mrm{b}^\top U_\mrm{b})^{-1} \tilde{b}(\alpha)$ and 
\begin{align*}
	C(\alpha) = \begin{pmatrix} c_\text{vol}(\alpha) \\ c_\text{peak}\end{pmatrix} = \begin{pmatrix}	\tilde{c}(\alpha) (U_\mrm{c}^\top P_\text{c})^{-1}U_\text{c}^\top \\ c_\mrm{peak}^\mrm{f} \end{pmatrix} V.
\end{align*}

\begin{figure}
	\centering
	\includegraphics{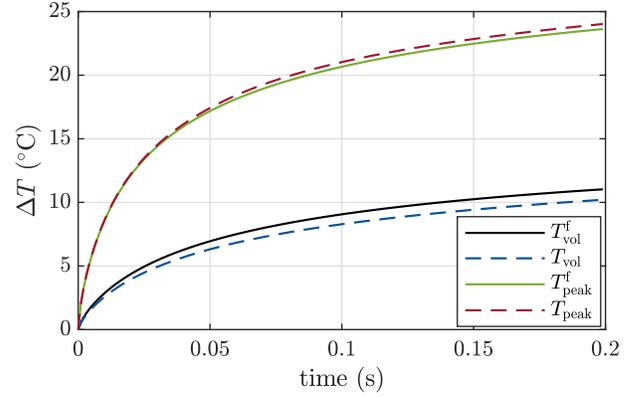}
	\caption{Peak and volume temperature of reduced and full order model with $\alpha = 0.38$ and input $u= 30\, \text{mW}$.}
	\label{fig:full_red}
\end{figure}

In this work, we use a reduction order $n = 6$ and a DEIM order $d = 3$ for ${\alpha} \in \R^1$   and $n = 7$ and $d = 3$ for ${\alpha} \in \R^2$. Both of these choices are high enough to render the relative peak and volume temperature error below 1\%, cf.\ \cite[Tables 5 and 7]{Schaller.2022a} and still allow us to perform model predictive control in real time \cite[Table 8]{Schaller.2022a}.   In Fig.~\ref{fig:full_red}, the peak and volume temperature of the full and reduced model for one parameter are shown. For illustration purposes, the extreme scenario $\alpha_\mrm{rpe} = 0.38$ is used, which is the lower bound of the parameter domain $\mathcal{D}$ and results in the largest errors over the domain $\mathcal{D}$. We note that this value is rather unlikely in experiments as it represents a deviation of the mean absorption prefactor by two standard deviations \cite[Table 3]{Schaller.2022a}. This value results in the largest error we observe over the whole parameter domain. The volume temperature is underestimated, whereas the peak temperature is slightly overestimated. The difference in the volume temperature influences the online parameter estimation, as will be shown in Sec.~\ref{sec:Comparison} and Sec.~\ref{sec:2p}.  

\section{Observer Designs} \label{sec:observerdesign}
\noindent We consider two different observer designs for joint state and parameter estimation. In this section, we recap the extended Kalman filter algorithm and the moving horizon estimator that we apply in the following sections. 

The continuous-time, reduced order model is discretized with the implicit Euler method to guarantee numerical stability at a sampling rate of $t_\mrm{s} = 0.001\,$s. This yields the discrete-time state-space model
\begin{align}
\label{eq:red_dModel}
\begin{split}
x_{k+1} &= A_\mrm{d} x_k + b_\mrm{d} (\alpha_k) u_k \\ 
y_k &= c_{\mrm{vol}}(\alpha_k) x_k,
\end{split}
\end{align}
with $A_\mrm{d}= (I-t_\mrm{s}A)^{-1}$ and $b_\mrm{d} (\alpha)=(I-t_\mrm{s}A)^{-1}t_\mrm{s}b(\alpha)$. 
To allow for joint state and parameter estimation, we extend the state-space model by adding the parameter $\alpha\in \R^p$ as an additional state, where $p = 1$, if only the absorption coefficient in the RPE is estimated, or $p = 2$ if both coefficients are estimated. As the coefficient is spot-dependent but time-invariant at one spot, we consider the constant parameter dynamics $\alpha_{k+1} = \alpha_{k}$. Furthermore, we assume that model~\eqref{eq:red_dModel} is subject to process noise $w_{k}$ and measurement noise $v_{k}$, which yields 
\begin{align}
	\begin{split}
		\bar{x}_{k+1} &= 
		\begin{pmatrix}
			x_{k+1} \\ \alpha_{k+1} \end{pmatrix} = f(x_k,\alpha_k,u_k) + w_k \\&= \begin{pmatrix}
			A_\mrm{d} & 0\\ 0 &1 
		\end{pmatrix} \bar{x}_k + \begin{pmatrix}
			b_\mrm{d}(\alpha_k)\\ 0
		\end{pmatrix} u_k +w_k, \\ 
		y_k &= g(x_k,\alpha_k)+v_k = \begin{pmatrix}
			c_{\mrm{vol}}(\alpha_k) & 0
		\end{pmatrix} \bar{x}_k + v_k.
	\end{split}
	\label{eq:ss_extend}
\end{align}

In the following, we set $\Vert v\Vert_{M} := \sqrt{v^TMv}$ for a vector $v\in \mathbb{R}^k$ and a positive definite matrix $M\in \mathbb{R}^{k\times k}$ with $k\in \mathbb{N}$. Furthermore, $|v|$ denotes the Euclidean norm of the vector $v\in \R^k$.

\subsection{Extended Kalman Filtering}\label{sec:ekf}
\noindent The extended Kalman filter is a well-known state estimator for nonlinear systems, see, e.g., \cite{Chui2017}. It is based on a successive linearization of the nonlinear model~\eqref{eq:ss_extend} at each time step $k$ where $w_{k}$ and $v_{k}$ are assumed to be uncorrelated and normally distributed.
The EKF algorithm can be divided into two steps. First, the a priori system state $\bar{x}_k^-$ and estimation error covariance matrix $P_k^- \in \R^{(n+p) \times (n+p)}$ are calculated from the previous estimates
\begin{align}
	\begin{split}
		\bar{x}_k^-&=f(\bar{x}_{k-1},u_{k-1})\\
		P_k^-&=A_{k-1}P_{k-1} A_{k-1}^\top+Q,
	\end{split}
\label{eq:EKF_pred}
\end{align}
where $Q \in \R^{(n+p)\times (n+p)}$ is the covariance matrix of the process noise and $A_{k-1} \in \R^{(n+p)\times (n+p)}$ is the Jacobian of $f$ with respect to $\bar{x}$, evaluated at $(\bar{x}_{k-1}, u_{k-1})$, i.e., 
\begin{align*}
	A_{k-1} &= \begin{pmatrix}
		A_\mrm{d} & \frac{\partial f(\bar{x}_{k-1},u_{k-1})}{\partial \alpha_{k-1}}|_{\bar{x}_{k-1},u_{k-1}}\\ 0 &1 
	\end{pmatrix}.
\end{align*}
Second, the estimation error covariance matrix $P_k$, the Kalman gain $H_k \in \R^{1 \times (n+p)}$ and the estimated state $\bar{x}_k$ are calculated as
\begin{align}
	\begin{split}
		H_k &= P_k^- c_k^\top (c_k P_k^- c_k^\top+R)^{-1}\\
		\bar{x}_k &= \bar{x}_k^- +H_k (y_k- g(\bar{x}_k^-))\\
		P_k&= (I_{n+1}-H_k c_k)P_k^-
	\end{split}
\label{eq:EKF_corr}
\end{align}
with the identity $I_{n+p} \in \R^{(n+p) \times (n+p)}$, the covariance of the measurement noise $R$, and the Jacobian of the output ${c_k = \frac{\partial g(\bar{x}_k)}{\partial \bar{x}_k}|_{\bar{x}_k^-}}$. The matrices $Q$ and $R$ are design parameters that weight the reliability of the model and the measurements. If the uncertainty about the measurements is high, $R$ is chosen large in relation to $Q$, and vice versa if the model uncertainty is large. 

\subsection{Moving Horizon Estimation}\label{sec:MHE}
\noindent Moving Horizon estimation is an optimization-based estimation technique for linear and nonlinear systems. Compared to EKF, MHE does not depend on a linearization of the system. In addition to the state, the MHE scheme can also be used to estimate the absorption parameter $\alpha$, in a similar manner as in EKF (i.e., by considering it as an additional state variable).
At each time instant $k$, a sequence of the extended states, i.e., the state $x$ and the parameter $p$, ${\bf \bar{x}}=\left(\bar{x}_{k-N|k},\ldots, \bar{x}_{k|k}\right)\in \mathbb{R}^{(N+1)(n+p)}$ is estimated for the past $N$ steps, where $N$ denotes the estimation horizon. Here, $\bar{x}_{i|k}$ denotes the (extended) state estimate for time $i$ (with $k-N\leq i\leq k$), estimated at the (current) time $k$. These estimated state sequence is obtained by solving an optimization problem, which takes into account the past $N$ measurements and inputs, as well as the system dynamics and possibly constraints (see, e.g., \cite{Rawlings.2017}). Given an initial guess $\chi\in \mathbb{R}^{n+p}$ for the extended state at time $k-N$, this optimization problem is given by
\begin{align}
	\label{eq:mhe.optimization}
	\begin{split}
	&\min_{{\bar{\bf{x}}}\in \mathbb{R}^{(N+1)(n+p)}}
	J(\bar{\bf x}) \\
	&\text{s.t. } \alpha_{i|k} \in [\alpha_\text{min},\; \alpha_\text{max}] \quad \forall \quad i = k-N...k,
	\end{split}
\end{align}
where the objective function $J$ is defined as
\begin{align}
	\begin{split}
	J(\bar{\bf{x}}):=&
	\left\|\bar{x}_{k-N|k}-\chi\right\|^2 _{P^{-1}} +
	\sum_{i=k-N}^{k} \Vert y_k - g(\bar{x}_{i|k})\Vert^2_{R^{-1}}\\
	&+ \sum_{i=k-N}^{k-1} 
	\left\|	\bar{x}_{i+1|k} - f(\bar{x}_{i|k}, u_k) \right\|^2_{Q^{-1}},
	\end{split}
	\label{eq:cost}
\end{align}
with symmetric positive definite weighting matrices $P,Q\in\R^{(n+p)\times (n+p)}$, weighting $R\in\R_{>0}$, and the lower and upper bounds $\alpha_\mrm{min} \in \R^p,\;\alpha_\mrm{max} \in \R^p$ with $p=1,2$ according to~\eqref{eq:paramdom}.  
The second term in the objective function $J$ penalizes the output-fitting error between the real measurement and the estimated output, i.e., the (estimated) measurement noise $v$. The third term penalizes the (estimated) process noise $w$. This means that the cost function~\eqref{eq:cost} trades off the confidence in the measurements and the model via a suitable choice of $Q$ and $R$ (similar to the EKF design, cf. Sec.~\ref{sec:tuningQRP}).
The optimal estimated state sequence at time $k$ is denoted by  $\bar{x}_k^* =(\bar{x}_{k-N|k}^*,\ldots, \bar{x}_{k|k}^*)$. 
Furthermore, the last elements of these optimal state sequence, i.e., $\bar{x}^*_{k|k}$,  
serve as the current state and parameter estimate at time $k$. 
Whenever $k<N$, we replace $N$ by $k$ in~\eqref{eq:mhe.optimization} and~\eqref{eq:cost}, i.e., we use the available amount of measurements to solve the optimization problem~\eqref{eq:mhe.optimization}.

The horizon length $N$, the prior $\chi$, as well as the weight matrices are design parameters which will be tuned in Sec.~\ref{sec:MHETuning}. In particular, we will take a closer look at the design of the prior $\chi$ and the weighting $P$. As pointed out in \cite{Valipour.2021,Lopez.2011}, a poor choice of $\chi$ and $P$ can result in an unnecessary large horizon length in order to achieve good estimation results. 
In view of real-time applicability of our methods, small horizons $N$ are desired, for which, in turn, the tuning parameters have a significant impact on the performance.

Hence, we use a common approach as suggested in \cite{Tenny.2002} and \cite{Qu2009} which is often used in practice and results in good estimation results. The EKF estimate of the error covariance matrix $P_i$ is used to update the prior weighting at each time k, i.e., $P = P_{k-N}$ with $P_{k-N}$ from~\eqref{eq:EKF_corr}. 

Furthermore, there exist different ways of updating the prior. Common choices are the so-called filtering update, where $\chi = \bar{x}_{k-N|k-N}$ and the smoothing update, where $\chi = \bar{x}_{k-N|k-1}$, see, e.g., \cite{Rakovic.2019}.      
The filtering update includes more (earlier) information in the MHE as measurements from $k-2N$ until $k-N$ were used to estimate $\bar{x}_{k-N|k-N}$. This can be interpreted as an extension of the horizon, \cite{Rakovic.2019}. A disadvantage is that the filtering update might need more time to recover from a bad initial prior. The smoothing update, on the other hand, includes measurements from $k-N-1$ until $k-1$, which means that the measurements from $k-N$ until $k-1$ are (indirectly) considered twice in the optimization problem \cite{Tenny.2002,Rakovic.2019}. 
 
\section{Comparison of EKF and MHE for one parameter}\label{sec:Comparison}
\noindent In this section, we perform an in-depth comparison of EKF and MHE. We first detail the tuning of both estimators in Sections \ref{sec:tuningQRP} and \ref{sec:MHETuning}. The tuning will be illustrated by showing exemplary and representative simulations, but the same conclusions regarding a suitable choice of tuning parameters also hold for different values of $\alpha$ and/or noise realizations. 
Then, in Sec.~\ref{sec:eval_simData}, we compare MHE and EKF in simulation, in particular the convergence speed, the sensitivity to normally distributed noise and relative errors between estimation and simulated data.
Afterwards, in Sec.~\ref{sec:eval_realData}, we discuss results with real measurement data from porcine eyes. Both observers are implemented in Matlab. For the MHE, CasADi \cite{Andersson.2019} and the nonlinear programming solver IPOPT \cite{Waechter.2006} were used. 

Before tuning the design parameters of EKF and MHE, we use the measurements from the parameter study, cf.\ Sec.~\ref{sec:modeling}, in order to estimate the variance of the measurement noise occurring in our experimental setup. At each of the 250 spots, the identified spot-dependent absorption coefficients in the RPE and choroid are used to perform open-loop simulations with a given constant input signal ($u = 30\; \text{mW}$). For each spot, we estimate the measurement noise $d_\mrm{noise}$ via the difference between the noise-free simulation $y_\mrm{sim}$ and the measured volume temperature $y_\mrm{meas}$, i.e., we set $d_\mrm{noise} = y_\mrm{meas}-y_\mrm{sim}$. Then, we calculate the variance of $d_\mrm{noise}$ for all spots individually and use its mean over all spots, denoted by $\mrm{var}_\mrm{meas} = 0.288\, ^\circ\text{C}$ to generate the noise that is added to the volume temperature in simulation.
\subsection{Tuning of $Q$, $R$ and $P_0$} \label{sec:tuningQRP}
\noindent In this section, we discuss the tuning of the covariance matrices $Q$ and $R$, as well as the initial error covariance matrix $P_0$ in the EKF. As the role and influence of the weighting matrices $Q$ and $R$ are the same for EKF and MHE, we deduce an appropriate tuning by means of only analyzing the EKF. 
Since in typical treatment situations, the initial temperature distribution is known precisely (no temperature increase has occurred before treatment), a very good initial prior is available for the states $x$. Conversely, a good initial prior for $\alpha$ is not necessarily available. For this reason, for the following tuning, we put particular focus on the convergence of the parameter estimates. 
The variance of the measurement noise, that was approximated as described above, could be used for $R$. However, the variance of the process noise $Q$ cannot be determined easily, since the parameter-dependent MOR error would need to be taken into account to this end. Instead, all states are weighted equally, i.e., we choose $Q = \mrm{diag}(0.01)$, which we found to give satisfactory state estimation results in combination with a suitable tuning of $R$ that is considered in the following.
\begin{figure}
	\centering
	\subfloat[\centering Volume temperature (gray) and estimated volume temperature with $R=10^2$ (blue), $R=10^3$ (green) and $R=10^4$ (red)]{{\includegraphics{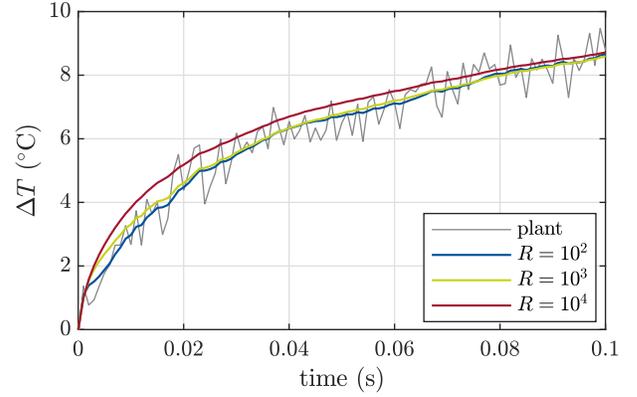}}\label{fig:diff_R_Tvol}}\\
	\subfloat[prefactor $\alpha$]{{\includegraphics{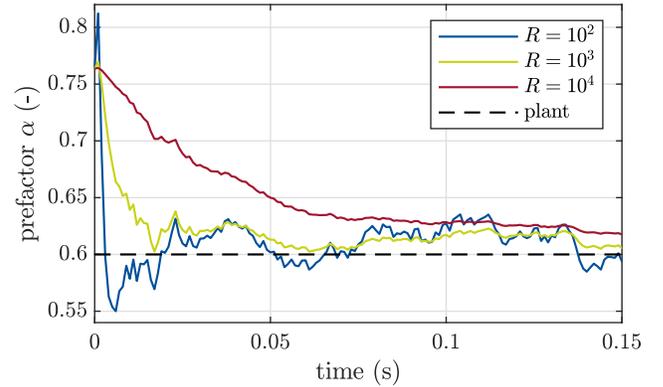}}{\label{fig:diff_R_alpha}}}
	\caption{Noisy simulation and EKF estimation with $u = 23\, \text{mW}$}
	\label{fig:diff_R}
\end{figure}

\begin{figure}
	\centering
	\includegraphics{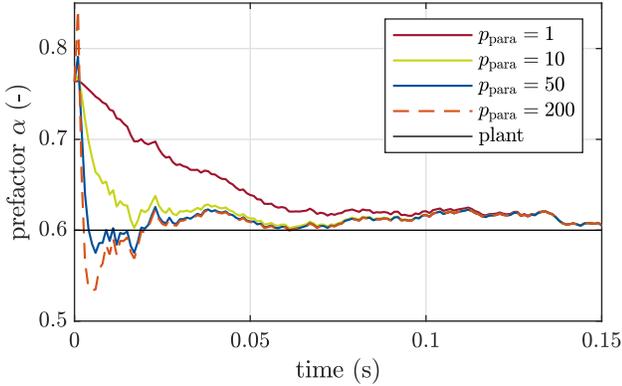}
	\caption{Estimation of prefactor $\alpha$ for different choices of $p_\mrm{para}$.}
	\label{fig:EKFP0}
\end{figure}

Figure~\ref{fig:diff_R} shows an exemplary simulated noisy trajectory of the output $T_\mrm{vol}$ together with the estimated outputs $\hat{T}_\mrm{vol}$ with different choices for $R$. If $R$ is chosen too large, the parameter converges very slowly, which results in an overestimation in the volume temperature as illustrated in Fig.~\ref{fig:diff_R_Tvol}. In contrast, if $R$ is too small, the parameter converges faster to a region around the actual parameter but is very sensitive to noise. Hence, $R = 10^3$ is chosen as a compromise. We point out that even for large values of $R$ and long simulation times, no convergence to the real parameter value is achieved due to the model mismatch between full order and reduced order model.
We use a block-diagonal initial prior weighting ${P_0 = \left(\begin{smallmatrix} P_\mrm{state} &\bs{0} \\ \bs{0} & p_\mrm{para} \end{smallmatrix} \right)}$, encoding the confidence on the initial state and parameter.
We choose $P_\mrm{state} = \text{diag}(0.01)$ and $p_\mrm{para}= 50$ to speed up the parameter estimation. The high value of $p_\mrm{para}$, in comparison to the elements in $P_\mrm{state}$, is chosen as the parameter value is initially unknown but the initial states are known quite precisely (no temperature increase before treatment, as discussed above). The choice of $P_0$ only influences the estimates at the beginning of the treatment. Nevertheless, the treatment duration is very short and hence, fast convergence is desired. The effect of different choices for $p_\mrm{para}$ are shown in Fig.~\ref{fig:EKFP0}. For $p_\mrm{para}= 1$, the parameter changes very slowly over time until all of the shown trajectories converges to the same values after approximately $0.13\,$s. For $p_\mrm{para} = 200$, the observer is tuned too aggressively and a considerable overshoot occurs. We found that $p_\mrm{para}= 50$ is a good compromise between convergence speed and a potential overshoot at the beginning independent of the choices for $\alpha$ and different noise realizations. 
\subsection{MHE specific tuning} \label{sec:MHETuning}
\noindent As discussed above (compare Section IV-B), additional tuning parameters for the MHE compared to EKF are the update of the prior $\chi$ and the weighting $P$. 
Concerning the former, we test the filtering and smoothing updates as described in Sec.~\ref{sec:MHE}. Concerning the latter, we test both a constant prior weighting $P$ as well as a time-varying prior weighting $P=P_{k-N}$, with $P_{k-N}$ from the EKF~\eqref{eq:EKF_corr}. For a constant prior weighting $P$, we use the block-diagonal structure as above and choose $P_\mrm{state}$ based on the solution of the algebraic Riccati equation. The reason for this choice is that for a fixed $\alpha$, this would be the stationary covariance matrix of the states $x$ in case of stationary process and measurement noise, \cite{Anderson.1979}. In order to compute $P_\mrm{state}$ from the algebraic Riccati equation, we use $b_d(\bar\alpha)$ and $c_\mrm{vol}(\bar\alpha)$ with $\bar\alpha=0.76$, which corresponds to the mean value of $\alpha$ over the considered parameter interval. Similar to above, we use $p_{para}=50$ in order to speed up the parameter estimation. 

\begin{figure}
	\centering
	\includegraphics{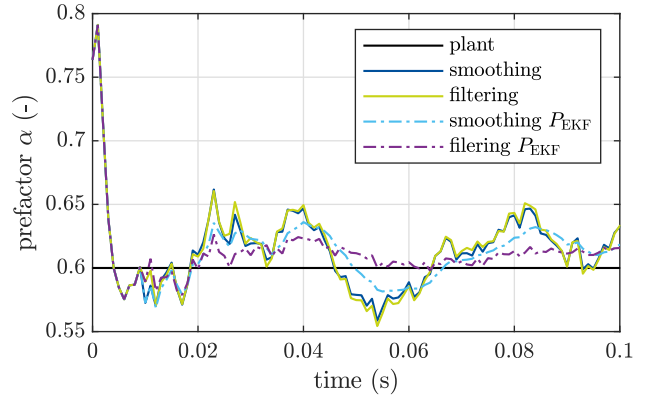}
	\caption{Estimation of parameter $\alpha$ with different choices of $\chi$ and $P$: smoothing update (blue) and filtering update (green) with constant $P$, smoothing update (blue,dashed) and filtering update (purple,dashed) with $P=P_{k-N}$ with $P_{k-N}$ from the EKF~\eqref{eq:EKF_corr}.}
	\label{fig:prior_alpha}
\end{figure}

In simulations without noise and model mismatch (reduced order model as plant), the state and parameter converge faster with the smoothing update for $Q = \mrm{diag}(0.01)$, $R = 10^3$, $N = 10$, and $P$ as described above. With output noise, we found that there is only a slight difference between the filtering and the smoothing update in the state and parameter estimation for the above choice of weighting matrices. However, when using the EKF update $P_{k-N}$ of the prior weight, the smoothing update is more sensitive to noise. In Fig.~\ref{fig:prior_alpha}, the parameter estimation with model mismatch (full order model) and output noise is shown for $\alpha = 0.6$. The filtering update with EKF prior weight has the smoothest trajectory. Similar results were obtained for different choices of $\alpha$ and noise realizations.   
 
\begin{figure}
	\centering
	\includegraphics{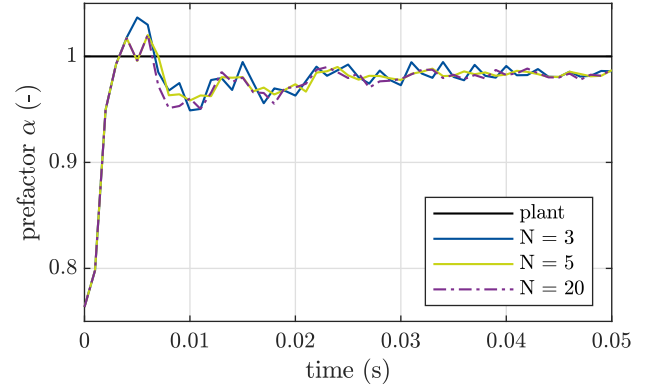}
	\caption{Estimation of parameter $\alpha$ with different horizon length $N = 3$ (blue), $N = 5$ (green), $N=20$ (dashed, purple).}
	\label{fig:horizon_alpha}
\end{figure}

Fig.~\ref{fig:horizon_alpha} depicts the influence of different horizon lengths $N$ for the filtering prior with EKF update for $\alpha = 1$. 
Whereas the convergence behavior of the parameter estimation seems to be almost independent of the horizon length $N$, the sensitivity to noise visibly decreases for increasing horizons. However, the relatively small horizon $N=5$ already provides a smooth estimation that is not strongly affected by noise, which is very favorable in view of future possible real-time application.

%
\subsection{Evaluation with simulated data}\label{sec:eval_simData}
\noindent In this section, we compare the performance of the MHE design and the EKF design in simulation. Given the findings of the previous sections, we consider the same weighting $Q = \mrm{diag}(0.01)$ and $R= 10^3$ in both observers, and for MHE we choose N=5, the filtering update for the prior and the EKF update for the prior weighting. We consider different values of $\alpha$, and for each of these values we performed simulations with 100 different noise realizations (normally distributed output noise with a variance of $0.288\, ^\circ\text{C}$) and a constant input of $30\, \text{mW}$. Such a constant heating process has been considered in previous open-loop control approaches \cite{Baade.2017} and will be sufficient for the identification of one parameter. In Sec.~\ref{sec:2p}, a more informative (time-varying) input will be needed in order to achieve good estimation and parameter identification results for two parameters.  

We evaluate the performance of both estimators in terms of the relative full state error $e_{\mrm{x},k} = \frac{|Vx_{\mrm{\{EKF,MHE\}},k}-x_k^\mrm{f}|}{|x_k^\mrm{f}|}$, the relative parameter error $e_{\alpha,k}=\frac{|\alpha_{\mrm{\{EKF,MHE\}},k}-\alpha_\mrm{sim}|}{\alpha_\mrm{sim}}$ and the relative errors in the peak and volume temperature $e_{\mrm{peak},k} = \frac{|c_\mrm{peak} x_{\mrm{\{EKF,MHE\}},k}-c_\mrm{peak}^\mrm{f}x_k^\mrm{f}|}{c_\mrm{peak}^\mrm{f}x_k^\mrm{f}}$ and $e_{\mrm{y},k} = \frac{|y_{\mrm{\{EKF,MHE\}},k}-y_k|}{y_k}$, respectively. The estimated output \newline
$y_{\mrm{\{EKF,MHE\}},k}= c_\mrm{vol}(\alpha_{\mrm{\{EKF,MHE\}},k})x_{\mrm{\{EKF,MHE\}},k}$ refers to the volume temperature. 

\begin{figure}
	\centering
	\includegraphics{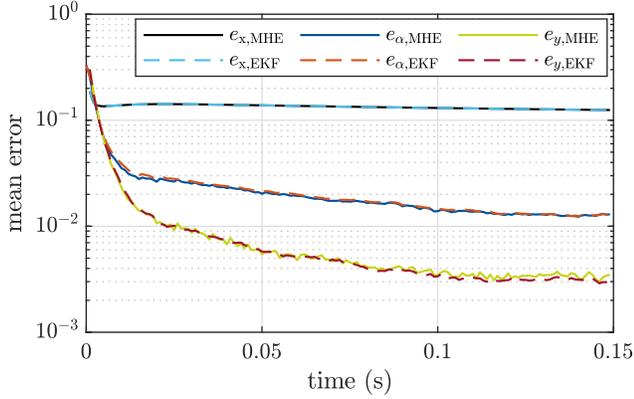}
	\caption{Mean error over all 100 noise sequences for ${\alpha = 1.14}$. }
	\label{fig:eSim_time}
\end{figure}
We define the average error at each time $t$ over all 100 noise sequences as $e_{*}(t) := \frac{1}{100} \sum_{s_n=1}^{100}e_{\star,k}(s_n)$ with $s_\mrm{n} = 1,\cdots,100$ (denoting the $s_\mrm{n}$-th noise realization), $* \in \{y,\alpha,x,\mrm{peak}\}$ and $t= k t_\mrm{s}$. In the same manner, we define the standard deviation of the relative error as $\sigma_*(t):= \text{std}(e_{*,k}(s_\mrm{n}))$. Figure~\ref{fig:eSim_time} shows $e_{*}(t)$ for $* \in \{y,\alpha,x\}$ for ${\alpha = 1.14}$. For the sake of clarity, we omit the peak temperature in Fig.~\ref{fig:eSim_time} (since the corresponding lines are very close to those of $e_{\alpha}(t)$) but analyze it later in Tab.~\ref{tab:compSim} (as it is our control variable). The error in the output is the lowest for both, MHE and EKF. The difference in $e_{\mrm{x}}(t)$ and $e_{\alpha}(t)$ between EKF and MHE is small and hard to distinguish. The larger error in $x$ and $\alpha$ can be explained by the model reduction error. The parameter is underestimated due to the model mismatch in the volume temperature as explained in Sec.~\ref{sec:modeling}, which also leads to an error in the states $Vx(t)$, where $V$ is the projection from the low dimensional space to the high dimensional space of the full model, cf.~\eqref{eq:red_c}.  

A more detailed comparison is given in Tab.~\ref{tab:compSim}, where we show the sum of the relative error, i.e., $\Sigma e_*:= \sum_{k=0}^{150} e_*(kt_\mrm{s})$ and the mean of the standard deviation of the error, i.e., $\bar{\sigma}_*:= \frac{1}{151} \sum_{k=0}^{150} \sigma_*(kt_\mrm{s})$.
One can see that both, the relative error and the mean of the standard deviation, are very similar for EKF and MHE. In particular, the sum of relative errors $\Sigma e_*$ is very similar; in some cases, MHE yields a slightly better $\Sigma e_*$ and EKF in other cases. For low values of $\alpha$, the MOR error is larger, hence, the estimated values are also less accurate. A slight difference can be seen in the standard deviations. The EKF seems to be slightly less sensitive to noise than the MHE because the standard deviation is marginally smaller in all cases for a horizon length of $N = 5$. For $N = 20$, the sum of the relative errors $\Sigma e_*$ and the standard derivatives $\bar{\sigma}_*$ are the same as the EKF values (for the considered decimal places as in Tab.~\ref{tab:compSim}). This means that for MHE, the sensitivity to noise becomes smaller with increasing horizon $N$ as expected.

\renewcommand{\arraystretch}{1.2}
\begin{table*}[t]
	\centering
\begin{minipage}[c]{13cm}
			\caption{Comparison of the sum of the errors $\Sigma e_*$ and the mean of the standard deviation $\bar{\sigma}_*$ for a simulation of $150\, \text{ms}$.}
	\begin{tabular}{c|c||c|c|c|c||c|c|c|c|}
		\multirow{2}{*}{$\alpha$}& & \multicolumn{4}{c||}{MHE} & \multicolumn{4}{c|}{EKF} \\
		& & $y$ & $\alpha$ & $ T_\mrm{peak}$& $x$ & $y$ & $\alpha$ & $T_\mrm{peak}$ & $x$\\
		\hline
		\multirow{2}{*}{1.14} & $\Sigma e_*$ & 1.74 &3.79 &2.87 &20.00 &1.71 &3.88 & 2.88 &20.00\\
		& $\bar{\sigma}_*$ & 0.0066 & 0.0097 &0.0050&0.001 & 0.0065& 0.0092&0.0049&0.001\\
		\hline
		\multirow{2}{*}{0.76} & $\Sigma e_*$ & 1.63& 1.84& 1.02& 3.48  & 1.56 & 1.75& 0.98& 3.47 \\
		& $\bar{\sigma}_*$ & 0.0085 & 0.0093& 0.0051 & 0.0020& 0.0081 & 0.0088& 0.0049&0.0019\\
		\hline
		\multirow{2}{*}{0.39} & $\Sigma e_*$ & 4.84& 23.30&13.30  & 26.86& 4.77& 23.18& 13.34&26.87\\
		& $\bar{\sigma}_*$ & 0.0168& 0.0300& 0.0154 &0.0021 & 0.0164& 0.0284& 0.0146&0.002\\
	\end{tabular}
\end{minipage}
	\label{tab:compSim}
\end{table*} 

\subsection{Evaluation with measurement data}\label{sec:eval_realData}
\begin{figure}
	\centering
	\subfloat[\centering volume temperature]{{\includegraphics{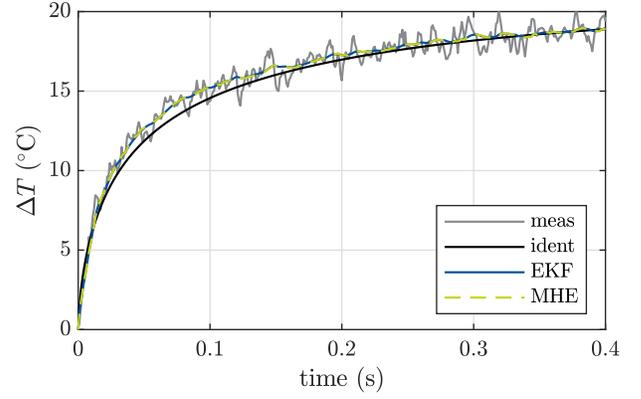}}\label{fig:meas11_Tvol}}\\
	\subfloat[prefactor $\alpha$]{{\includegraphics{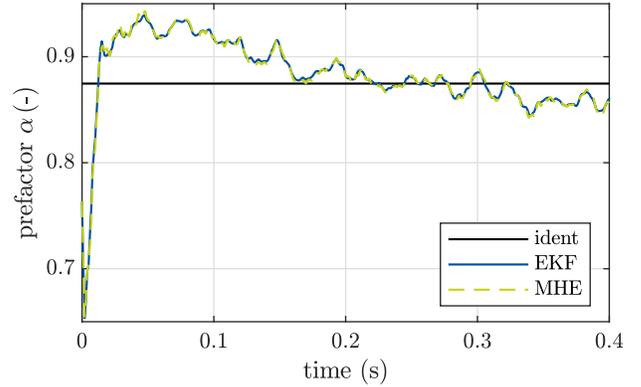}}{\label{fig:meas11_alpha}}}
	\caption{Irradiated porcine eye explant with $u = 30\, \text{mW}$: measurement (grey), simulation with offline identified $\alpha_\text{ident}$ (black), EKF estimation (blue) and MHE estimation (green, dashed).}
	\label{fig:meas11}
\end{figure}
\begin{figure}
	\centering
	\includegraphics{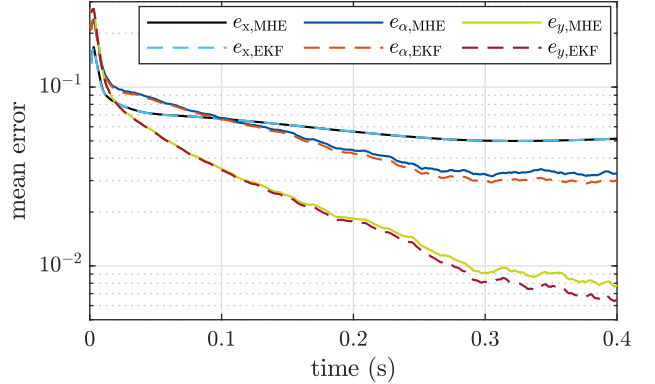}
	\caption{Error between full order simulation and the EKF and MHE estimations, respectively.}
	\label{fig:error_ekf_mhe_meas}
\end{figure}

\noindent After intensive tuning of EKF and MHE, we use the measurements from our case study for further evaluation of both estimation techniques on experimental data. A horizon length of $N = 5$ is used for MHE, as discussed in Sec.~\ref{sec:MHETuning}. Figure~\ref{fig:meas11} illustrates results for one out of the 250 measurements with $u = 30\, \text{mW}$. As the absorption coefficient $\mu$ and hence, the parameter $\alpha$ can not be measured in experiments, we use the value $\alpha_\mrm{ident}$ we obtained from the case study (by offline identification, i.e., taking all output measurements into account, cf. Sec.~\ref{sec:modeling} and \cite{Schaller.2022a}) as a reference value for the parameter estimation. It can be seen that there is a discrepancy between the simulation with the full order model (black curve) and the measurements (gray curve). This explains the fact that the (online) estimated parameter is initially higher than $\alpha_\mrm{ident}$. However, after approximately $180\,\text{ms}$, the estimated parameter stays in a region around $\alpha_\mrm{ident}$. Furthermore, it can be seen that both estimators are hard to distinguish, which confirms the findings with simulated data. To allow for further comparison, we utilize the offline identified prefactor $\alpha_\text{ident}$ for simulations with the full order model. 
Figure~\ref{fig:error_ekf_mhe_meas} shows the relative errors $e_{*}(t)$ between the simulations of the full model with $\alpha_\mrm{ident}$ and the estimated values of the EKF and MHE with the corresponding measurement data. Here, $e_{*}(t)$ is the mean of all 250 measurement spots with a spot-dependent $\alpha$ (and not the mean over different noise realizations for a constant $\alpha$ as in Fig.~\ref{fig:eSim_time}), i.e., ${e_{*}(t) := \frac{1}{250} \sum_{s_n=1}^{250}e_{\star,k}(s_n)}$ with $s_n = 1,...,250$. We point out that this error is computed with respect to offline simulations due to a shortcoming of real (state and parameter) data, and hence is only a surrogate for the real error. Nevertheless, these simulations serve as a reference to compare MHE and EKF. Again, there is only a small difference between MHE and EKF, with EKF having a slightly smaller error. This observation is also supported by Tab.~\ref{tab:compMeas}, which shows the sum of the relative error $\Sigma e_*$ and the average standard deviation of the error $\bar{\sigma}_*$ as defined in the previous subsection, but for a measurement length of 400\,ms (i.e., $k=400$ instead of $k=150$). 
\begin{table*}[t]
	\centering
	\caption{Comparison of the sum of the errors $\Sigma e_*$ and the mean of the standard deviation $\bar{\sigma}_*$ for measurement data (first $400\, \text{ms}$)}
	\begin{tabular}{c||c|c|c|c||c|c|c|c|}
		 & \multicolumn{4}{c||}{MHE} & \multicolumn{4}{c|}{EKF} \\
		 & $y$ & $\alpha$ & $ T_\mrm{peak}$& $x$ & $y$ & $\alpha$ & $T_\mrm{peak}$ & $x$\\
		\hline
		 $\Sigma e_*$ & 11.84& 22.22 &11.75 &24.27 & 11.52 & 21.37 &11.46 &24.26\\
		 $\bar{\sigma}_*$ & 0.021 &0.077 &0.033 & 0.044 &0.019 & 0.064& 0.030 & 0.044\\
	\end{tabular}
	\label{tab:compMeas}
\end{table*} 

\section{Estimation with two parameters}\label{sec:2p}
\noindent In this section, we show results for the joint state and parameter estimation when considering the two absorption coefficients in the RPE and the choroid as two independent parameters. As pointed out in Sec.~\ref{sec:modeling}, choosing the absorption coefficient in the choroid to be constant leads to a slightly different temperature increase. Therefore, we investigate also the performance of EKF and MHE for two independent absorption coefficients. We consider a reduced model of order seven (cf. Sec.~\ref{sec:modeling}) and extend the state by the two-dimensional parameter vector $\alpha = [\alpha_\mrm{rpe},\, \alpha_\mrm{ch}]^\top$, as described in Sec.~\ref{sec:observerdesign}. First, we describe differences from the one parameter case and analyze the behavior in simulations for different absorption coefficients by means of $\Sigma e_*$ and $\bar{\sigma}_*$. Afterwards, we show results with measurement data in more detail. 

\subsection{Evaluation with simulated data}\label{sec:sim_2p}
\begin{figure}
	\centering
	\includegraphics{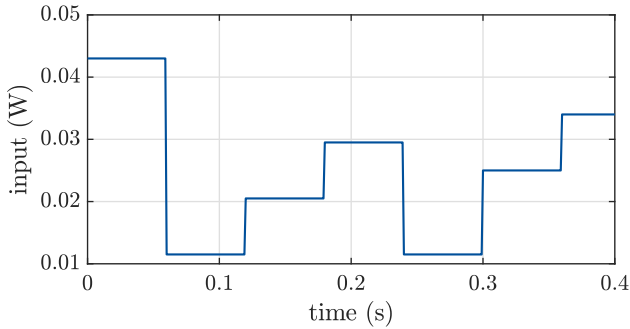}
	\caption{Time varying input signal }
	\label{fig:fancyu}
\end{figure}

\noindent For the estimation of two parameters, we found that a constant input is not exciting enough to ensure parameter convergence. In fact, convergence can in general not be guaranteed in a joint parameter and state estimation (for EKF) and biased estimations are possible as pointed out in \cite{Ljung1979}. In our case, we found that the obtained bias in the absorption coefficients when using a constant input signal depends on various factors, in particular on $Q$, the magnitude of the input signal and the specific parameter values. In order to improve the parameter estimation, we instead use a more informative, time-varying input signal as depicted in Fig.~\ref{fig:fancyu}. 
For realistic parameter configurations of $\alpha_\mathrm{rpe}$ and $\alpha_\mrm{ch}$ in simulation, we choose three different offline identified combinations, i.e., one with low absorption, one close to the average value and one with high absorption (in the RPE), according to Tab.~\ref{tab:compSim_2p}. 

We found that a slightly different tuning of $Q$ than in the one parameter case (1p-case) is more suitable, i.e., $q_\mrm{para}=\mrm{diag}(0.005,0.001)$, where $q_\mrm{para}$ is the part of the covariance matrix $Q$ that belongs to the parameters, comparable with $p_\mrm{para}$ in $P_0$ (cf. Sec.~\ref{sec:tuningQRP}). Moreover, we tuned $p_\mrm{para}$ such that intensive overshoots of $\alpha_\mrm{ch}$ in the (unconstrained) EKF are avoided, i.e., $p_\mrm{para} = \mrm{diag}(50,20)$. As in the 1p-case, a larger horizon $N$ for MHE did not show a significant improvement in estimation, therefore we apply $N=5$. Moreover, the EKF update $P_{k-N}$ and the filtering update are used for MHE. 

In the following, we consider the full order model for simulations of the plant and add normally distributed noise with a variance of $0.228^\circ$C to the simulated output (as for the 1p-case). We performed simulations with 100 different noise realizations and the time-varying input according to Fig.~\ref{fig:fancyu}. Table~\ref{tab:compSim_2p} shows the sum of errors $\Sigma e_*$ and the mean of the standard deviation $\bar{\sigma}_*$ for two parameters. 
For large and average (RPE) absorption coefficients, EKF and MHE result in similar values for both $\Sigma e_*$ and $\bar{\sigma}_*$, with EKF resulting in slightly smaller values.
Exceptions are $\Sigma e_x$ and $\bar{\sigma}_x$, where the MHE performs slightly better. For low (RPE) absorption, the MHE estimation errors are lower than for EKF, except for the estimation of $\alpha_\mrm{ch}$. In general, $\alpha_\mrm{ch}$ is hard to estimate correctly as can be seen in the rather large error sum $\Sigma e_{\alpha,\mrm{ch}}$ and the deviation $\bar{\sigma}_{\alpha,\mrm{ch}}$ for all considered different absorption coefficients. A possible explanation for this is that the sensitivity with respect to the choroid absorption is smaller than w.r.t.\ its counterpart in the RPE \cite{Schaller.2022a}, making it harder to identify differences in the choroid absorption parameter from the outputs. Furthermore, we observed that for the chosen input signal, approximately 0.1\,s are needed for both observers to converge into a neighborhood of the true values, independent of the specific parameter values for the chosen input signal. 
\begin{table*}[t]
		\caption{Comparison of the sum of the errors $\Sigma e_*$ and the mean of the standard deviation $\bar{\sigma}_*$ for a simulation with a time varying u and $400\, \text{ms}$ for two parameter.}
	\centering
	\begin{tabular}{c|c|c||c|c|c|c|c||c|c|c|c|c|}
		\multirow{2}{*}{$\alpha_\mrm{rpe}$}& \multirow{2}{*}{$\alpha_\mrm{ch}$}& & \multicolumn{5}{c||}{MHE} & \multicolumn{5}{c|}{EKF} \\
		& & & $y$ & $\alpha_\mrm{rpe}$ & $\alpha_\mrm{ch}$&  $ T_\mrm{peak}$& $x$ & $y$ & $\alpha_\mrm{rpe}$ & $\alpha_\mrm{ch}$& $T_\mrm{peak}$ & $x$\\
		\hline
		\multirow{2}{*}{1.12} & \multirow{2}{*}{0.07} & $\Sigma e_*$ & 2.78 & 11.38& 80.30& 4.72& 53.95& 2.63& 10.34 & 78.69& 4.49&54.02\\
		& & $\bar{\sigma}_*$ & 0.0050& 0.0157& 0.0742& 0.0037& 0.0023& 0.0047& 0.0143& 0.0743&0.047 &0.0025\\
		\hline
		\multirow{2}{*}{0.76} & \multirow{2}{*}{0.09} & $\Sigma e_*$ & 3.38 & 8.53& 21.25& 1.96& 6.92& 3.15& 7.75& 20.31&1.70 &6.71\\
		& & $\bar{\sigma}_*$ & 0.0065& 0.0159& 0.0392& 0.0038& 0.0039& 0.0060& 0.0145&0.0385 & 0.0033&0.0035\\
		\hline
		\multirow{2}{*}{0.39} & \multirow{2}{*}{0.1} & $\Sigma e_*$ & 6.12 & 46.34& 52.88& 31.46& 67.62&6.12 & 62.87& 46.84& 33.89& 68.47\\
		& & $\bar{\sigma}_*$ & 0.0102& 0.0499& 0.0501& 0.0102& 0.0020& 0.0100& 0.0491& 0.0526& 0.0096&0.0029\\
	\end{tabular}
	\label{tab:compSim_2p}
\end{table*} 
\subsection{Evaluation with measurement data}\label{sec:meas_2p}
\begin{figure}
	\centering
	\subfloat[\centering volume temperature]{{\includegraphics{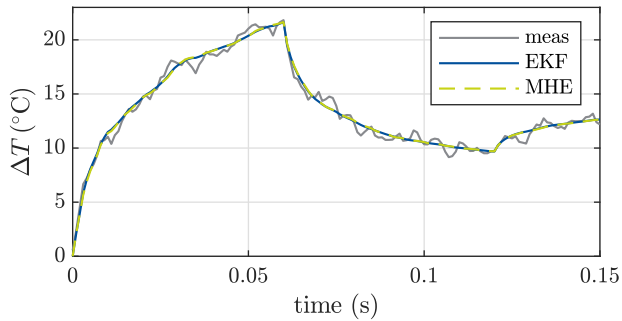}}\label{fig:Tvol_meas_2p}}\\
	\subfloat[prefactor $\alpha_\mrm{rpe}$]{{\includegraphics{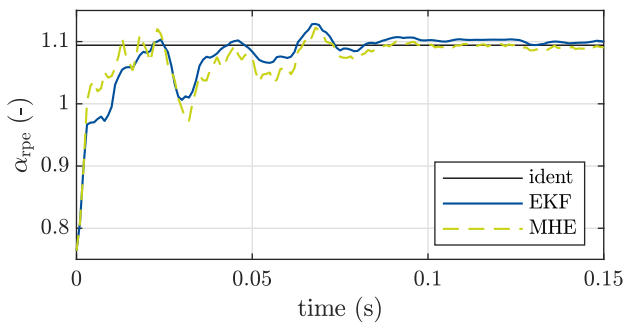}}{\label{fig:alpha1_meas_2p}}}\\
	\subfloat[prefactor $\alpha_\mrm{ch}$]{{\includegraphics{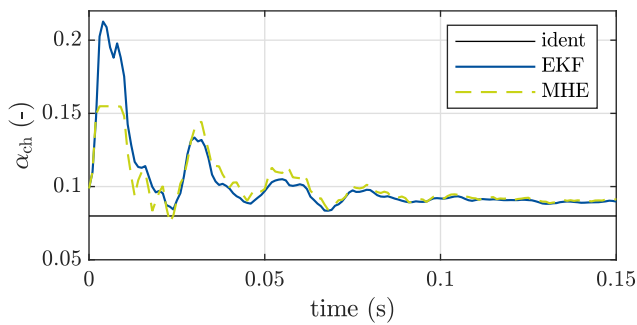}}{\label{fig:alpha2_meas_2p}}}
	\caption{Irradiated porcine eye explant: measurement (grey), offline identified $\alpha_\text{ident}$ (black), EKF estimation (blue) and MHE estimation (green, dashed).}
	\label{fig:meas_2p}
\end{figure}

\noindent After tuning both estimators for two parameters in simulation, we will have a closer look at the estimation of two parameters with experimental data. As in the 1p-case, we use the offline identified absorption coefficients $\bs{\alpha}_\mrm{ident} \in \R^2$ that we obtained from the case study as a reference for the parameter estimates. Figure~\ref{fig:meas_2p} shows one of the 250 measurements with the time varying input depicted in Fig.~\ref{fig:fancyu}. Whereas the estimates of the volume temperature are barely distinguishable, the parameter estimates for both estimators differ especially for the first 100\,ms. The overshoot of $\alpha_\mrm{ch}$ in the EKF for the first 20\,ms is quite strong, which corresponds to a slower initial increase of $\alpha_\mrm{rpe}$. For MHE, $\alpha_\mrm{rpe}$ increases faster as a result of the active constraints on $\alpha_\mrm{ch}$, and therefore, the overshoot is smaller. Here, for the estimation of two parameters, the results in the parameter estimation differ between EKF and MHE due to the active constraints. This behavior was not observed in the 1p-case, where the results are alike for EKF and MHE. However, after a settling time, both estimators perform, once again, very similarly. 

\begin{figure}
	\centering
	\includegraphics{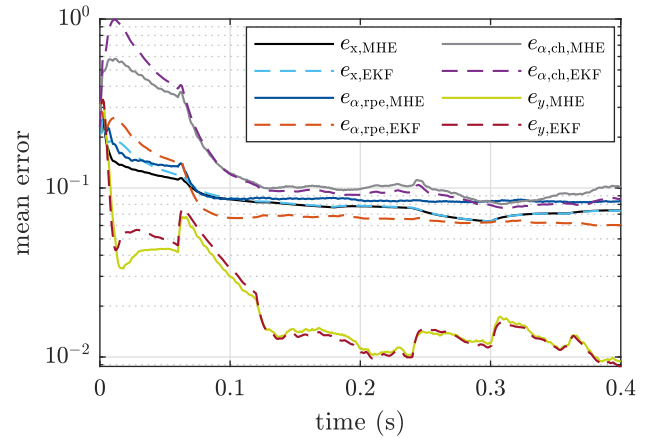}
	\caption{Mean error between the full order simulation and the EKF and MHE, respectively, with two parameters. }
	\label{fig:emeas_time_2p}
\end{figure}

As for simulated data, the estimation in average is more accurate for $\alpha_\mrm{rpe}$ than for $\alpha_\mrm{ch}$. This is also indicated by the mean (relative) errors over all 250 measurements in Fig.~\ref{fig:emeas_time_2p}. As in the 1p-case, we utilize $\bs{\alpha}_\mrm{ident}$ for simulations with the full order model and compare those to the MHE and EKF results, respectively. It can be seen that the relative error $e_{\alpha,\mrm{ch}}$ is larger than $e_{\alpha,\mrm{rpe}}$, $e_{x}$ or $e_{y}$, except for a short time interval around 0.3\,s where $e_{\alpha,\mrm{ch}}$ is slightly smaller than $e_{\alpha,\mrm{rpe}}$ for MHE. Due to the constraints included in the MHE formulation, MHE initially performs better than the EKF. After approximately 100\,ms, the parameter estimation is better for EKF than for MHE for both absorption coefficients. This can be explained as follows. We used the 250 measurements also in our case study to compute the average absorption and the standard deviation. Note that for the bounds in MHE, we consider two times the standard deviation from the average value (cf. Sec.~\ref{sec:MHE}), i.e., 95\,\% of the absorption coefficients are inside the bounds. 
In Fig.~\ref{fig:emeas_time_2p}, we also considered the outliers, i.e., $\bs{\alpha}_\mrm{ident} \notin [\alpha_\mrm{min},\,\alpha_\mrm{max}]$ in~\eqref{eq:mhe.optimization}, which results in a worse parameter estimation for MHE than for EKF (as $\bs{\alpha}_\mrm{ident}$ is unattainable if it is outside the bounds of ${\alpha}$ in~\eqref{eq:mhe.optimization}). However, if the constraints on $\alpha$ are not included in the MHE formulation, the advantage of MHE over EKF in the first 100\,ms vanishes and both observers perform similarly whereas the EKF performs still slightly better than MHE. If we exclude the outliers in Fig.~\ref{fig:emeas_time_2p} and the bounds on $\alpha$ are included in the MHE formulation, the MHE performs, again, better at the beginning (due to constraints). After some settling time, both estimators are hard to distinguish. Furthermore, the errors are slightly smaller for the state and parameter estimates for both observers if the outliers are excluded. 

In Tab.~\ref{tab:compMeas_2p_woOutlier}, the sum of the errors $\Sigma e_*$ and the mean of the standard deviation $\bar{\sigma}_*$ are shown for EKF and MHE. The outliers are excluded in the table to have a fair comparison between MHE and EKF if $\bs{\alpha}_\mrm{ident}$ is inside the bounds. Due to the active parameter constraints that affect the estimation at the beginning (cf. Fig.~\ref{fig:meas_2p} and Fig.~\ref{fig:emeas_time_2p}), the sum of the errors $\Sigma e_*$ is smaller for MHE. The smaller overshoot in the estimation for MHE is also reflected in the mean of the standard deviations $\bar{\sigma}_*$ which are also smaller for MHE than for EKF. 

\begin{table*}[t]
	\centering
\caption{Comparison of the sum of the errors $\Sigma e_*$ and the mean of the standard deviation $\bar{\sigma}_*$ for measurement data (first $400\, \text{ms}$) without outliers.}
	\begin{tabular}{c||c|c|c|c|c||c|c|c|c|c|}
		& \multicolumn{5}{c||}{MHE} & \multicolumn{5}{c|}{EKF} \\
		& $y$ & $\alpha_\mrm{rpe}$ & $\alpha_\mrm{ch}$& $ T_\mrm{peak}$& $x$ & $y$ & $\alpha_\mrm{rpe}$ & $\alpha_\mrm{ch}$ &  $T_\mrm{peak}$ & $x$\\
		\hline
		$\Sigma e_*$ & 9.92 & 24.13& 57.78& 8.20&28.74 & 10.43& 27.66& 67.33& 8.60& 30.32\\
		$\bar{\sigma}_*$ & 0.0153 & 0.0415& 0.1023& 0.0172& 0.0348& 0.0157& 0.0477& 0.1194& 0.0178& 0.0356\\
	\end{tabular}
	\label{tab:compMeas_2p_woOutlier}
\end{table*} 
  
The estimation performance can further be enhanced by using different input signals. For example, one option would be to use sinusoidal input signals with different frequencies, which are such that the two different absorption coefficients mostly affect the input-output behavior. Using experiment design techniques \cite{Hjalmarsson.2007,Pintelon.2012} to design tailored input signals for an improved estimation is an interesting subject of future work. Furthermore, in future closed-loop experiments, it needs to be ensured that the applied input is sufficiently exciting to allow for parameter estimation. Alternatively, a separate short identification phase at each spot prior to treatment could be used. This will be subject of future work.

\section{Conclusion}\label{sec:conclusion}
\noindent We have compared two different estimators, MHE and EKF, for joint parameter and state estimation in the context of retinal laser therapies. Two different scenarios were considered, the estimation of one unknown absorption coefficient and two unknown absorption coefficients. We evaluated MHE and EKF for different absorption coefficients and noise realizations in simulation and with measurement data in 1\,kHz. Overall, we found that both estimators for both scenarios perform well in view of our application. In the case of one unknown parameter, there is only a slight difference between EKF and MHE. The estimation of two unknown parameters turned out to be more difficult than one parameter leading to differences between EKF and MHE. Especially at the beginning of the estimation, MHE benefits from active parameter constraints whereas the EKF shows large overshoots. This could be crucial in closed-loop control without an additional identification phase and where MHE might outperform EKF. A detailed evaluation of closed-loop estimation also with regard to a sufficiently exciting input (in the case of two unknown parameters) is part of future work. One aspect we did not discuss in this paper is the real-time capability of MHE, as it is computationally more demanding than EKF. Considering this aspect, for real-time closed-loop (model predictive) control in 1\,kHz, fast sub-optimal schemes for MHE such as \cite{Kuehl.2011,Schiller.2021} might by beneficial. This is also part of future work.

\section*{Acknowledgment}
\noindent The authors would like to thank Julian D. Schiller for helpful discussions.

\bibliographystyle{IEEEtran} 
\bibliography{references}

\end{document}